# A non-local model for the description of twinning in polycrystalline materials in the context of infinitesimal strains: application to a magnesium alloy

Charles Mareau[1] and Hamidreza Abdolvand[2]

[1] Arts et Métiers Institute of Technology, LAMPA, HESAM Université, F-49035 Angers, France
[2] Department of Mechanical and Materials Engineering, Western University, London, Ontario, N6A 5B9, Canada

A polycrystalline plasticity model, which incorporates the contribution of deformation twinning, is proposed. For this purpose, each material point is treated as a composite material consisting of a parent constituent and multiple twin variants. In the constitutive equations, the twin volume fractions and their spatial gradients are treated as external state variables to account for the contribution of twin boundaries to free energy. The set of constitutive relations is implemented in a spectral solver, which allows solving the differential equations resulting from equilibrium and compatibility conditions. The proposed model is then used to investigate the behavior of a AZ31 magnesium alloy. For the investigated loading conditions, the mechanical behavior is mostly controlled by the joint contribution of basal slip and tensile twinning. Also, according to the numerical results, the development of crystallographic texture, morphological texture and internal stresses is consistent with the experimental observations of the literature.

**Keywords:** microstructures, twinning, constitutive behaviour, crystal plasticity, spectral method

## 1 Introduction

For metallic alloys with a hexagonal close packed (hcp) structure, the mechanical behavior is impacted by the competition between deformation twinning and crystallographic slip. Indeed, for such materials, plastic deformation is partly accommodated by mechanical twinning because of the limited number of easy slip systems.

In contrast with crystallographic slip, which is a rather progressive deformation mechanism, twinning results in significant and rapid microstructural transformations, even at moderate strains. For instance, many experimental studies have shown that, depending on loading conditions, texture development is affected by twinning in the early stages of deformation (Brown et al. 2005; Jiang et al. 2007; Xu et al. 2009). Also, due to the resistance opposed by surrounding grains, the growth of deformation twins is associated with the development of an internal stress field. Neutron diffraction techniques have allowed investigating the evolution of this internal stress field through the evaluation of the resulting lattice strains. Such techniques have been used to study load partition between twin and parent grains in magnesium (Wu et al. 2008; Clausen et al. 2008) and zirconium (Xu et al. 2008b) alloys. In the recent years, some alternative techniques with a higher spatial resolution have been used to evaluate internal stresses. For instance, Arul Kumar et al. (2018) used micro-Laue X-ray diffraction techniques to estimate the stress state in the vicinity of a twin in magnesium. They concluded that the formation of a twin divides the parent grain into two non-interacting domains. Nervo et al. (2016) used diffraction contrast tomography to investigate the twinning behavior of a titanium alloy. Their results suggest that prismatic $\langle a \rangle$ slip favors the formation and clustering of twins. Twin variant selection in titanium has been studied with high resolution electron backscatter diffraction techniques by Guo et al. (2017). They found that variant selection is controlled by the local stress state. Three dimensional X-ray diffraction (3D-XRD) techniques (Poulsen et al. 2001) have allowed measuring the positions, orientations, volume fractions and elastic strains of multiple grains during the *in-situ* deformation of a magnesium alloy (Aydıner et al. 2009; Abdolvand et al. 2015). The corresponding dataset has been used to identify twin-parent pairs and evaluate the grain volume-averaged stress tensors. For a given twin-parent pair, Aydıner et al. (2009) and Abdolvand et al. (2015) found that the average normal stress acting on the habit plane is almost the same for the twin and parent grains. However, for a twin grain, the average tangential stress acting on the twin habit plane along the twin direction is lower than that of the corresponding parent grain. Recently, *in situ* EBSD techniques (Gussev et al. 2018) have been used to investigate the impact of pre-existing twins and dislocations on twin nucleation in $\alpha$-uranium (Grilli et al. 2020b).

Different strategies have been explored to include the contribution of twinning in the context of polycrystalline plasticity. Mean-field polycrystalline models have been widely used with either rate-independent (Agnew et al. 2006; Xu et al. 2008a; Clausen et al. 2008) or rate-dependent (Proust et al. 2007; Mareau and Daymond 2011; Zhang and Joshi 2012; Wang et al. 2018) crystal plasticity-based constitutive models. Though some solutions have been proposed (Cherkaoui 2003; Juan et al. 2014), the interactions between parent grains and twin domains are generally evaluated only in an averaged manner with the aforementioned approaches. Also, by definition, mean-field polycrystalline models ignore the role of intragranular stresses on the deformation behavior. Full-field polycrystalline models rely on either the finite element method (Kalidindi 2001; Abdolvand et al. 2011; Staroselsky and Anand 2003; Chang and Kochmann 2015; Cheng and Ghosh 2015; Ardeljan et al. 2016; Paramatmuni and Dunne 2020; Indurkar et al. 2020; Ravaji and Joshi 2021) or the spectral method (Mareau and Daymond 2016; Paramatmuni and Kanjarla 2019) to evaluate the intragranular stress and strain gradients resulting from twinning. Since the associated constitutive models often use a pseudo-slip formulation, the morphological texture evolution resulting from the formation of twins is not always considered. To circumvent these limitations, some procedures have been proposed to insert twinned domains in a polycrystalline microstructure (Ardeljan et al. 2015; Cheng and Ghosh 2017). However, such procedures require an *a priori* knowledge of the twin morphology and ignore the contribution of interfacial energy. The phase-field method provides a convenient thermodynamic framework to consider the increase of surface energy resulting from the formation of twin boundaries (Clayton and Knap 2011; Kondo et al. 2014; Liu et al. 2018; Grilli et al. 2020a). In the context of twinning, the phase-field method treats the twin volume fraction and its spatial gradient as state variables, which allows tracking twin boundaries without any complex numerical procedure.

In the present work, a non-local constitutive model, which incorporates the contribution of twinning to plastic deformation, is proposed. Constitutive relations are developed in the context of crystal plasticity. Also, the proposed model is non-local in the sense that the spatial gradients of the twin volume fractions are included in the list of state variables to account for the contribution of twin boundaries to free energy. The model is then implemented within a spectral solver, which allows solving the differential equations resulting from equilibrium and compatibility conditions. In contrast with previous implementations (Mareau and Daymond 2016; Paramatmuni and Kanjarla 2019), non-locality provides a way of circumventing the difficulties associated with plastic strain localization resulting from twinning-induced softening.

The present paper is organized as follows. The constitutive equations are provided in the first part. In the second part, the proposed model is used to investigate the impact of deformation twinning on the behavior of a AZ31 magnesium alloy. The numerical results allow discussing several aspects of deformation twinning, including twin variant selection, texture development and internal stress evolution.

## 2 Model description

### 2.1 Composite material assumption

As illustrated in Figure 1, the present approach, which is developed in the infinitesimal strain framework, treats each material point as a composite material to deal with the growth of twinned domains within a polycrystalline volume element. Each material point thus consists of $n + 1$ constituents: one parent constituent (with volume fraction $\phi_m$) and $n$ child constituents (with volume fractions $\phi_{c=1,n}$), which correspond to the $n$ twin variants. For any material point, the volume fractions should satisfy the condition

$$\sum_c \phi_c + \phi_m = 1. \tag{1}$$

**Figure 1** Each material point is treated as a composite material consisting of a parent constituent and multiple child constituents corresponding to the different twin variants.

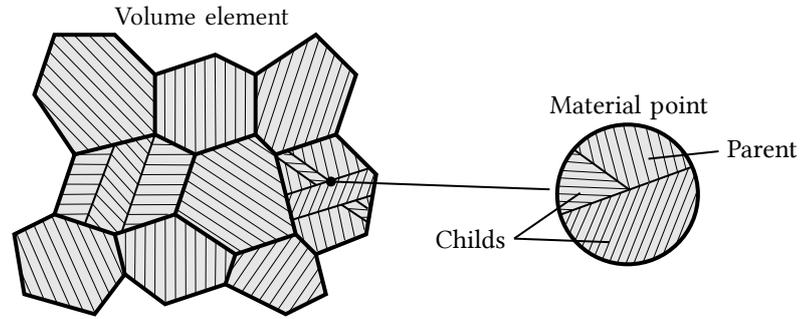

According to the averaging relations of homogenization theory, the total strain tensor is

$$\boldsymbol{\varepsilon} = \sum_c \phi_c \boldsymbol{\varepsilon}_c + \phi_m \boldsymbol{\varepsilon}_m. \tag{2}$$

In the following, the stress state is assumed to be uniform within the different constituents during a deformation process, that is:

$$\boldsymbol{\sigma}_c = \boldsymbol{\sigma}_m = \boldsymbol{\sigma}, \tag{3}$$

where $\boldsymbol{\sigma}$ is the stress tensor. As a result of the composite material assumption, deformation twinning is treated as a progressive process. Indeed, a material point can be either fully untwinned (i.e. $\phi_m = 1$), fully twinned (i.e. $\phi_m = 0$) or in a intermediate situation (i.e. $0 < \phi_m < 1$). Also, while the stress state is uniform within the different constituents of a material point, the stress state is not identical for the parent and twinned domains of a given grain, which is composed of an infinite number of material points.

### 2.2 Equilibrium equations

The development of the constitutive model for twinning relies on the introduction of an extended set of state variables. In contrast with the classical theory, which uses the infinitesimal strain tensor $\boldsymbol{\varepsilon}$ as the sole external state variable[1], the volume fractions $\phi_c$ associated with the different twin variants and the corresponding spatial gradients $\boldsymbol{\nabla}\phi_c$ are also treated as external state variables. The consequence is that the power developed by internal forces includes the contributions resulting from the evolution of the strain field and twin volume fractions. Specifically, the density of power developed by internal forces $p_i$ is given by (Nguyen 2015)

$$p_i = \boldsymbol{\sigma} : \dot{\boldsymbol{\varepsilon}} + \sum_c \xi_c \dot{\phi}_c + \sum_c \boldsymbol{\eta}_c \cdot \boldsymbol{\nabla}\dot{\phi}_c, \tag{4}$$

where $\boldsymbol{\sigma}$ is the Cauchy stress tensor, $\xi_c$ is the scalar microstress power-conjugate to the volume fraction rate $\dot{\phi}_c$ and $\boldsymbol{\eta}_c$ is the vector microstress power-conjugate to the spatial gradient $\boldsymbol{\nabla}\dot{\phi}_c$. The composite material assumption allows reformulating the density of power developed by internal forces as

$$p_i = \sum_c \phi_c \boldsymbol{\sigma} : \dot{\boldsymbol{\varepsilon}}_c + \phi_m \boldsymbol{\sigma} : \dot{\boldsymbol{\varepsilon}}_m + \sum_c \dot{\phi}_c \boldsymbol{\sigma} : (\boldsymbol{\varepsilon}_c - \boldsymbol{\varepsilon}_m) + \sum_c \xi_c \dot{\phi}_c + \sum_c \boldsymbol{\eta}_c \cdot \boldsymbol{\nabla}\dot{\phi}_c. \tag{5}$$

According to the above expression, the power developed by internal forces is spent to deform the individual constituents and to expand/contract twinned domains.

The evolution of the strain field and twin volume fractions is governed by equilibrium equations which can be derived from the application of an extended principle of virtual power, see (Frémond and Nedjar 1996) in the context of damage or (Cermelli and Gurtin 2002) in the context of crystal plasticity. The direct consequence of this assumption is that the classical equilibrium equations of continuum mechanics are supplemented with the additional equilibrium equations[2]

$$\boldsymbol{\sigma} \cdot \boldsymbol{\nabla} = \boldsymbol{0} \text{ and } \boldsymbol{\sigma} = \boldsymbol{\sigma}^T, \tag{6}$$

$$\xi_c - \boldsymbol{\nabla} \cdot \boldsymbol{\eta}_c = 0, \ \forall c. \tag{7}$$

---

[1] In a thermomechanical context, one would treat temperature as an additional external state variable.

In the literature, the later equation is often referred to as the microforce balance equation (Cermelli and Gurtin 2002).

## 2.3 State equations

For a material point, the free energy density $a$ is decomposed according to:

$$a = \sum_c \phi_c a_c + \phi_m a_m + \sum_c \gamma_{c/m}, \tag{8}$$

where $a_m$ and $a_c$ are the bulk contributions of parent and child constituents while $\gamma_{c/m}$ is the surface contribution associated with a parent/twin interface. For the construction of constitutive relations, the total strain tensor is decomposed into elastic (superscript e), plastic (superscript p) and transformation (superscript tw) contributions. In the context of infinitesimal transformations, one obtains $\boldsymbol{\varepsilon}_c = \boldsymbol{\varepsilon}_c^e + \boldsymbol{\varepsilon}_c^p + \boldsymbol{\varepsilon}_c^{tw}$ and $\boldsymbol{\varepsilon}_m = \boldsymbol{\varepsilon}_m^e + \boldsymbol{\varepsilon}_m^p$. It is worth mentioning that, for a child constituent, the transformation strain is calculated from the characteristic shear strain $\gamma^{tw}$, which is constant, with $\boldsymbol{\varepsilon}_c^{tw} = \text{sym}(\boldsymbol{t}_c^\alpha \otimes \boldsymbol{k}_c^\alpha)\gamma^{tw}$, where $\boldsymbol{t}_c$ is the twinning direction and $\boldsymbol{k}_c$ is the twin plane normal associated to the corresponding twinning system. Also, within the framework of crystal plasticity, the plastic strain tensors are given by

$$\boldsymbol{\varepsilon}_c^p = \sum_\alpha \text{sym}(\boldsymbol{m}_c^\alpha \otimes \boldsymbol{n}_c^\alpha)\gamma_c^\alpha \quad \text{and} \quad \boldsymbol{\varepsilon}_m^p = \sum_\alpha \text{sym}(\boldsymbol{m}_m^\alpha \otimes \boldsymbol{n}_m^\alpha)\gamma_m^\alpha, \tag{9}$$

where $\boldsymbol{m}^\alpha$, $\boldsymbol{n}^\alpha$ and $\gamma^\alpha$ are respectively the slip direction, the slip plane normal and the plastic shear strain of the $\alpha$th slip system for either the parent or child constituents (subscript $m$ or $c$). The contribution of crystallographic slip to the plastic strain tensor of a child constituent is considered here to describe the potential stress relaxation associated with the possible existence of easy slip systems within twins (Louca et al. 2021).

In the following, the bulk contributions to free energy are defined as follows:

$$a_c = \frac{1}{2}\boldsymbol{\varepsilon}_c^e : \mathbb{C}_c : \boldsymbol{\varepsilon}_c^e + \frac{1}{2}\sum_\alpha \varrho_c^\alpha \sum_\beta H^{\alpha\beta} \varrho_c^\beta, \tag{10}$$

$$a_m = \frac{1}{2}\boldsymbol{\varepsilon}_m^e : \mathbb{C}_m : \boldsymbol{\varepsilon}_m^e + \frac{1}{2}\sum_\alpha \varrho_m^\alpha \sum_\beta H^{\alpha\beta} \varrho_m^\beta. \tag{11}$$

In the above equations, the first term represents the elastic strain energy while the second term corresponds to defect energy. The stiffness tensors associated with the parent or child constituents are denoted by $\mathbb{C}_m$ and $\mathbb{C}_c$. Also, $\varrho_m^\alpha$ and $\varrho_c^\alpha$ are the isotropic hardening variables associated with the $\alpha$th slip system. The possible interactions between different slip systems are described with the matrix $H$.

For the interfacial energy, the suggestion of Liu et al. (2018), which includes both local and non-local contributions, is adopted:

$$\gamma_{c/m} = \frac{1}{2}\boldsymbol{\nabla}\phi_c \cdot \boldsymbol{K}_c \cdot \boldsymbol{\nabla}\phi_c + A\phi_c\left(1 - \phi_c\right). \tag{12}$$

The local contribution is minimum for either $\phi_c = 0$ or $\phi_c = 1$, which correspond to the equilibrium states of a fully untwinned or fully twinned material point. A simple choice for $\boldsymbol{K}_c$ consists in writing:

$$\boldsymbol{K}_c = B\boldsymbol{k}_c \otimes \boldsymbol{k}_c + C\left(\boldsymbol{1} - \boldsymbol{k}_c \otimes \boldsymbol{k}_c\right). \tag{13}$$

---

[2] For simplicity, the contributions of volume forces and inertia effects are excluded.

The anisotropic gradient contribution allows considering the fact that the interfacial energy along the twin boundary is smaller than that of the twin tip. The surface energy associated with the twin/parent interfaces, as well as the width of the twin/parent interfaces, are controlled by the $A$, $B$ and $C$ material parameters[3].

For the development of constitutive relations, it is convenient to define the thermodynamic forces associated with the different state variables. First, when no viscous contribution to the stress state is considered, the stress tensors $\sigma_c$ and $\sigma_m$ are obtained from:

$$\sigma_c = \frac{\partial a_c}{\partial \varepsilon_c} = \mathbb{C}_c : \varepsilon_c^e, \tag{14}$$

$$\sigma_m = \frac{\partial a_m}{\partial \varepsilon_m} = \mathbb{C}_m : \varepsilon_m^e. \tag{15}$$

In a similar fashion, if the microstress vector $\eta_c$ has a purely energetic origin (i.e. it does not depend on $\nabla \dot\phi_c$), the corresponding state equation is

$$\eta_c = \frac{\partial a}{\partial \nabla \phi_c} = K_c \cdot \nabla \phi_c. \tag{16}$$

The resolved shear stresses $\tau_c^\alpha$ and $\tau_m^\alpha$, which act on the different slip systems, are evaluated from the projections of the corresponding stress tensors:

$$\tau_c^\alpha = -\frac{\partial a_c}{\partial \gamma_c^\alpha} = m_c^\alpha \cdot \sigma_c \cdot n_c^\alpha = m_c^\alpha \cdot \sigma \cdot n_c^\alpha, \tag{17}$$

$$\tau_m^\alpha = -\frac{\partial a_m}{\partial \gamma_m^\alpha} = m_m^\alpha \cdot \sigma_m \cdot n_m^\alpha = m_m^\alpha \cdot \sigma \cdot n_m^\alpha. \tag{18}$$

The critical resolved shear stresses (CRSS) $r_c^\alpha$ and $r_m^\alpha$ are connected to the corresponding isotropic hardening variables according to:

$$r_c^\alpha = \frac{\partial a_c}{\partial \varrho_c^\alpha} = \sum_\beta H^{\alpha\beta} \varrho_c^\beta \quad \text{and} \quad r_m^\alpha = \frac{\partial a_m}{\partial \varrho_m^\alpha} = \sum_\beta H^{\alpha\beta} \varrho_m^\beta. \tag{19}$$

Finally, the thermodynamic force associated with the twin volume fraction $\phi_c$, which is denoted by $\zeta_c$, is given by:

$$\zeta_c = \frac{\partial a}{\partial \phi_c} = (a_c - a_m) + A(1 - 2\phi_c). \tag{20}$$

From the definition of the thermodynamic forces, the evolution rate for the specific free energy is

$$\dot a = \sum_c \left[\phi_c \left(\sigma_c : \dot\varepsilon_c - \sum_\alpha \tau_c^\alpha \dot\gamma_c^\alpha - r_c^\alpha \dot\varrho_c^\alpha\right) + \zeta_c \dot\phi_c + \eta_c \cdot \nabla \dot\phi_c\right] + \phi_m \left(\sigma_m : \dot\varepsilon_m - \sum_\alpha \tau_m^\alpha \dot\gamma_m^\alpha - r_m^\alpha \dot\varrho_m^\alpha\right). \tag{21}$$

## 2.4 Evolution equations

The second law of thermodynamics allows defining the density of power being dissipated into heat. In a purely mechanical context, this quantity, which is denoted by $d$, is obtained from the difference between the power developed by internal forces and the evolution rate for free energy (Gurtin et al. 2010), that is $d = p_i - \dot a$. Combining Equation (21) and Equation (5), the expression of the dissipation source becomes

$$d = \sum_c \left[\phi_c \left(\sum_\alpha \tau_c^\alpha \dot\gamma_c^\alpha - r_c^\alpha \dot\varrho_c^\alpha\right) + \pi_c \dot\phi_c\right] + \phi_m \left(\sum_\alpha \tau_m^\alpha \dot\gamma_m^\alpha - r_m^\alpha \dot\varrho_m^\alpha\right) \tag{22}$$

where the dissipative force $\pi_c = \xi_c - \zeta_c + \sigma : (\varepsilon_c - \varepsilon_m)$, which governs the evolution of the twin volume fraction, is introduced. Combining the equilibrium Equation (7) and the state Equation (20) and Equation (16) leads to the complete expression

$$\pi_c = \nabla \cdot (K_c \cdot \nabla \phi_c) - A(1 - 2\phi_c) + (a_m - a_c) + \sigma : (\varepsilon_c - \varepsilon_m). \tag{23}$$

---

[3] The surface energy density is equal to $\pi\sqrt{AB/32}$ (respectively $\pi\sqrt{AC/32}$) for a twin/parent interface that is parallel (respectively perpendicular) to the twin plane.

According to the above expression, the dissipative force driving the evolution of the twin volume fraction depends not only on the current stress state, but also on the free energy difference between the parent and child constituents and the twin volume fraction in the neighborhood of the material point of interest. This aspect is consistent with the results of Paramatmuni et al. (2021), who observed that twin nucleation is governed by the stored energy density. The term $A(1 - 2\phi_c)$ can be interpreted as the resistance to deformation twinning. The material parameter $A$ therefore represents a barrier to the initiation of deformation twinning. Also, when the twin volume fraction increases, the resistance to deformation twinning decreases. The progression of twinning is therefore accompanied by a softening phenomenon that promotes plastic strain localization. While a local version of the model would be mesh dependent (because of the stress softening produced by twinning), the non-local contribution allows circumventing the difficulties associated with strain localization. Indeed, when the twin volume fraction excessively increases within a single material point, the non-local contribution favours the growth of the twin domain toward neighbouring material points.

In the context of standard materials (Nguyen 2000), the evolution equations, which connect the flux variables to the dissipative forces, are obtained from a dissipation potential $\varphi$. For the proposed model, the dissipation potential is decomposed according to

$$\varphi = \sum_c \phi_c \varphi_c + \phi_m \varphi_m + \sum_c \kappa_{c/m}. \tag{24}$$

The contributions $\varphi_c$ and $\varphi_m$ of the child and parent constituents to the dissipation potential are given by

$$\varphi_c = \sum_\alpha \frac{K}{N+1}\left(\frac{\langle |\tau_c^\alpha| - r_c^\alpha \rangle}{K}\right)^{N+1} \quad \text{and} \quad \varphi_m = \sum_\alpha \frac{K}{N+1}\left(\frac{\langle |\tau_m^\alpha| - r_m^\alpha \rangle}{K}\right)^{N+1}. \tag{25}$$

In the above equations, $K$ and $N$ are viscosity parameters, which control the resistance to plastic flow and the strain-rate sensitivity. Also, for the contribution of twinning to the dissipation potential, the following choice is adopted:

$$\kappa_{c/m} = \phi_m \frac{L}{M+1}\left(\frac{\langle \pi_c \rangle}{\gamma^{\text{tw}} L}\right)^{M+1} + \phi_c \frac{L}{M+1}\left(\frac{\langle -\pi_c \rangle}{\gamma^{\text{tw}} L}\right)^{M+1}, \tag{26}$$

where $L$ and $M$ are viscosity parameters whose role resembles that of $K$ and $N$. It is emphasized that the contribution of both twinning and detwinning to the dissipation potential have been included in the above expression.

The evolution equations for the different flux variables can be obtained from the differentiation of the dissipation potential with respect to the corresponding dissipative forces. Specifically, the shear strain rates associated with the different slip systems are given by

$$\dot{\gamma}_c^\alpha = \frac{\partial \varphi_c}{\partial \tau_c^\alpha} = \left(\frac{\langle |\tau_c^\alpha| - r_c^\alpha \rangle}{K}\right)^N \text{sign}(\tau_c^\alpha) \quad \text{and} \quad \dot{\gamma}_m^\alpha = \frac{\partial \varphi_m}{\partial \tau_m^\alpha} = \left(\frac{\langle |\tau_m^\alpha| - r_m^\alpha \rangle}{K}\right)^N \text{sign}(\tau_m^\alpha). \tag{27}$$

The evolution of the isotropic hardening variables is controlled by

$$\dot{\varrho}_c^\alpha = -\frac{\partial \varphi_c}{\partial r_c^\alpha} = \left(\frac{\langle |\tau_c^\alpha| - r_c^\alpha \rangle}{K}\right)^N = |\dot{\gamma}_c^\alpha| \quad \text{and} \quad \dot{\varrho}_m^\alpha = -\frac{\partial \varphi_m}{\partial r_m^\alpha} = \left(\frac{\langle |\tau_m^\alpha| - r_m^\alpha \rangle}{K}\right)^N = |\dot{\gamma}_m^\alpha|. \tag{28}$$

Finally, the evolution equation for the twin volume fraction associated with the $c$th twin variant is

$$\dot{\phi}_c = \frac{\partial \varphi}{\partial \pi_c} = \frac{\phi_m}{\gamma^{\text{tw}}}\left(\frac{\langle \pi_c \rangle}{\gamma^{\text{tw}} L}\right)^M - \frac{\phi_c}{\gamma^{\text{tw}}}\left(\frac{\langle -\pi_c \rangle}{\gamma^{\text{tw}} L}\right)^M. \tag{29}$$

It is worth mentioning that the twin volume fraction is not allowed to increase (respectively decrease) when the material point is fully twinned (respectively fully detwinned).

### 2.5 Texture evolution

In the present context, texture evolution is controlled by two different phenomena: (i) lattice rotation due to crystallographic slip and (ii) twinning. To consider the contribution of lattice rotation, the initial orientation of a (child or parent) constituent is specificied with a set of three Euler angles that can be updated with the Bunge's formula (Bunge 1968) from the lattice spin tensor. The lattice spin tensor of a (child or parent) constituent is obtained from the difference between the total spin tensor $\dot{\omega}$, which is the skew-symmetric part of the velocity gradient tensor, and the plastic spin tensor, which is calculated from the plastic shear strain rates. The lattice spin tensor of a child or parent constituent ($\dot{\omega}_c^e$ or $\dot{\omega}_m^e$) is therefore calculated according to

$$\dot{\omega}_c^e = \dot{\omega} - \sum_\alpha \text{skw}(\boldsymbol{m}_c^\alpha \otimes \boldsymbol{n}_c^\alpha)\dot{\gamma}_c^\alpha \quad \text{and} \quad \dot{\omega}_m^e = \dot{\omega} - \sum_\alpha \text{skw}(\boldsymbol{m}_m^\alpha \otimes \boldsymbol{n}_m^\alpha)\dot{\gamma}_m^\alpha. \tag{30}$$

The contribution of twinning to texture evolution is naturally accounted for in the present framework. Indeed, a child constituent does not contribute to the initial crystallographic texture since the corresponding volume fraction is zero. When a twinning system is activated, the volume fraction increases and the contribution of a child constituent to the global crystallographic texture of the volume element progressively increases. It is worth mentioning that the above approach considers that the child-parent orientation relationship is not impacted by plastic strains, which would be inappropriate in the context of finite strains.

## 3 Application to a AZ31 magnesium alloy

In this section, the proposed model is used to investigate the impact of deformation twinning on the behavior of a AZ31 magnesium alloy, for which some experimental results have been recently obtained by Louca et al. (2021). Specifically, Louca et al. (2021) have extracted some specimens from a rolled AZ31 magnesium plate. The specimens have been annealed to obtain an equiaxed microstructure, with an average grain size of 50 μm. Three-dimensional synchrotron X-ray diffraction techniques (Poulsen et al. 2001) have then been used to determine the position, volume, crystallographic orientation and elastic strains of individual grains during the *in situ* deformation of the polycrystalline AZ31 magnesium alloy. Based on these results, the twin-parent pairs have been identified and the corresponding stress tensors have been evaluated.

For the application of the present model, the field equations resulting from equilibrium and compatibility conditions are solved numerically with the spectral method (Moulinec and Suquet 1998; Eisenlohr et al. 2013). As shown in Equation (23), the dissipative force driving the evolution of the twin volume fraction involves the divergence of the microstress vector. In the present work, the finite difference method is used for the evaluation of the non-local contribution $\nabla \cdot (\boldsymbol{K}_c \cdot \nabla \phi_c)$ to this dissipative force. The numerical implementation of the proposed model is briefly described in Section A.

### 3.1 Microstructure generation

The volume element used for the application of the spectral method has been obtained from a weighted Voronoi tessellation with fifty seeds. For each seed, the position and weight are respectively given by the center-of-mass and the cube root of the volume of the corresponding grain. The resulting volume element ($150 \times 150 \times 150\,\mu m^3$), which is shown in Figure 2, has been discretized into $128^3$ voxels for the application of the spectral method. The size of a voxel is about 1 μm. The impact of resolution on numerical results in discussed in Section B. Due to the rolling process, most grains have their c-axis aligned with the normal direction of the plate.

### 3.2 Loading conditions

In the present work, two different loading paths have been selected to reproduce the experiment by Louca et al. (2021). The first loading path corresponds to a uniaxial compression test along the rolling direction with an axial strain rate of $5.5 \times 10^{-5}\,s^{-1}$ up to an axial strain of $-0.5\,\%$. The second loading path, which uses the same strain rate, corresponds to a Bauschinger test. A uniaxial compression test is first carried out along the rolling direction up to an axial strain of $-0.31\,\%$. The volume element is then unloaded and finally deformed under uniaxial tension up to an axial strain of $0.13\,\%$. While the selected loading conditions do not allow a full assessment of the proposed model, the corresponding numerical results can be compared to the experimental dataset of Louca et al. (2021) to see whether the description of twinning is consistent with experimental observations or not.

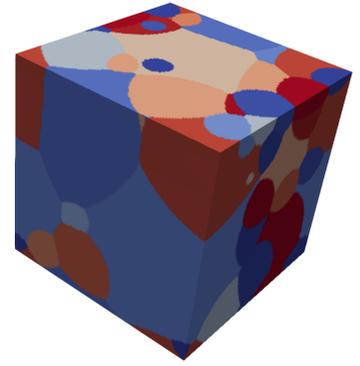

**Figure 2**  Numerical microstructure used for the application of the spectral method. Each color corresponds to a grain.

### 3.3 Material parameters

The material parameters used to model the mechanical behavior of the AZ31 magnesium alloy are listed in Table 1. The single crystal elastic constants have been taken from the work of

| Elasticity | | | | | Crystallographic slip | | | | |
|---|---|---|---|---|---|---|---|---|---|
| $\mathbb{C}_{11}$ | $\mathbb{C}_{33}$ | $\mathbb{C}_{12}$ | $\mathbb{C}_{13}$ | $\mathbb{C}_{44}$ | $K$ | $N$ | $r_{\text{Bas}}$ | $r_{\text{Pri}}$ | $r_{\text{Pyr}}$ |
| 59.8 GPa | 61.7 GPa | 23.2 GPa | 21.7 GPa | 16.4 GPa | 2.0 MPa | 20 | 5.6 MPa | 28.0 MPa | 46.7 MPa |

| Twinning | | | | | Hardening |
|---|---|---|---|---|---|
| $L$ | $M$ | $A$ | $B$ | $C$ | $H^{\alpha\beta}$ |
| 2.0 MPa | 20 | 3.6 MPa | 0.015 µN | 0.15 µN | 1000 MPa |

**Table 1**  Material parameters used for the numerical simulation of the AZ31 magnesium alloy. $r_{\text{Bas}}$, $r_{\text{Pri}}$ and $r_{\text{Pyr}}$ are the initial CRSS associated with basal (Bas), first order prismatic (Pri) and second order pyramidal (Pyr) slip systems.

Simmons and Wang (1971). Four plastic deformation modes are considered: basal (Bas), first order prismatic (Pri) and second order pyramidal (Pyr) slip systems and tensile twinning (TTw). The initial values for the CRSS associated with the different deformation modes and the hardening parameters have been adjusted to reproduce the macroscopic stress-strain behavior under uniaxial compression (see Figure 3 and Figure 9). For simplicity, no distinction is made between latent and self-hardening (i.e. $H^{\alpha\beta} = H^{\alpha\alpha}$). It is worth mentioning that the ratios between the CRSS of the different slip modes correspond to those given by Clausen et al. (2008): $r_{\text{Pri}}/r_{\text{Bas}} = 5$, $r_{\text{Pyr}}/r_{\text{Bas}} = 8.3$, $r_{\text{TTw}}/r_{\text{Bas}} = 5$[4]. The values of the $B$ and $C$ parameters control the energy density for a twin/parent interface that is either parallel or perpendicular to the twin plane. In the present work, the value of $B$ has been selected to obtain a coherent twin boundary energy density of $0.13\,\text{J/m}^2$ (Pei et al. 2017). Also, the ratio $C/B$ has been fixed to 10 to favour twin propagation over twin expansion.

### 3.4 Results and discussion

#### 3.4.1 Uniaxial compression

The experimental and numerical stress-strain curves obtained under uniaxial compression along the rolling direction are plotted in Figure 3. The relative activities[5] of the different slip and

---

[4] Though the CRSS for twinning $r_{\text{TTw}}$ does not appear explicitly in the description of the constitutive model, it is actually given by the ratio $A/\gamma^{\text{tw}}$.

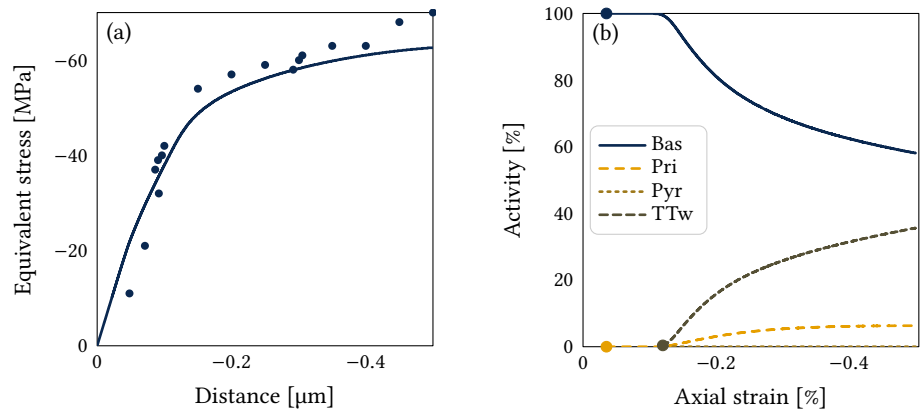

**Figure 3**  (a) Experimental and numerical stress-strain curves obtained for a uniaxial compression test along the rolling direction. Dots correspond to the experimental data of Louca et al. (2021) while solid lines correspond to numerical data. (b) Relative activities of the different inelastic deformation modes. The first activation of a given type of slip or twinning system is represented with a dot.

twinning deformation modes are also presented. Basal slip systems are first activated (around $-16$ MPa), which corresponds to the initial yielding. The activation of twinning systems (around $-41$ MPa) and prismatic slip systems (around $-43$ MPa) coincides with the transition to the plastic regime. Pyramidal slip systems are never activated during the deformation process.

The local twin volume fraction, which is given by $1 - \phi_m$, is plotted in Figure 4. Though the prescribed strain is rather small, twinned domains can clearly be identified at the end of the compression test. The strong morphological texture evolution due to twinning is therefore correctly reproduced. It is worth mentioning that the proposed model does not require any rule to define how nucleation and propagation take place. The formation of twinned domains is the direct consequence of the application of equilibrium, compatibility and constitutive equations.

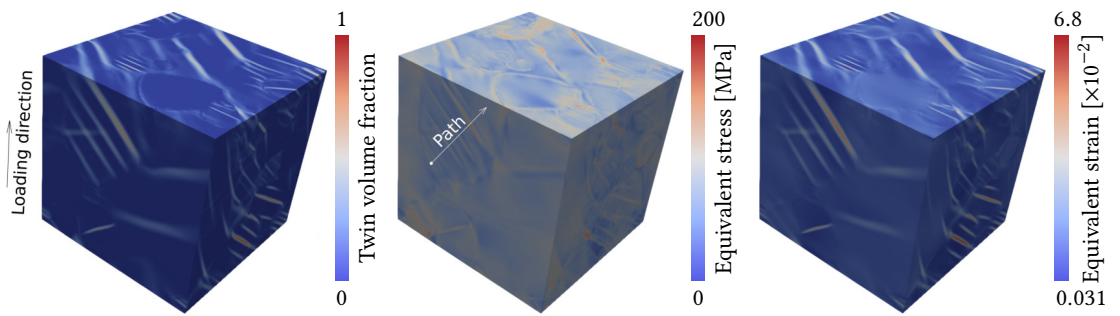

**Figure 4**  End of a uniaxial compression test.

Due to twin variant selection, most twins are oriented with an angle of approximately ±45° with respect to the loading direction. To further investigate how twin variant selection takes place, the twin volume fraction of each variant is plotted as a function of the Schmid factor in Figure 5. Two different options are explored for the evaluation of the Schmid factor. The global Schmid factor is calculated for each twin variant from the orientation of the corresponding system. This factor provides some information regarding the orientation of a twinning system with respect to the macroscopic stress state. The local Schmid factor is evaluated for each twin variant from the grain-averaged stress tensor of the corresponding to parent grain. It therefore considers the impact of the intergranular internal stresses resulting from inhomogeneous plastic deformation. Immediately after yielding (i.e. $-0.16\%$), no matter how the Schmid factor is evaluated, the twin volume fraction is generally higher for favourably oriented twin variants. At the end of the compression test (i.e. $-0.5\%$), the maximal twin volume fraction is obtained for some grains for which the local Schmid factor is not maximal, which is a consequence of the local stress relaxation resulting from deformation twinning. A more quantitative approach of interpreting the local/global Schmid factors consists of considering the 52 twin variants with the most important activity (i.e. with a volume fraction higher than 1 % at the end of the compression test). Table 2 shows the average Schmid factor and the corresponding standard deviation that have been calculated for these most active variants. According to the results, the standard deviation is much higher for the global Schmid factor, which indicates that the propensity of a given twin variant to be activated is better described with the local Schmid factor than the global one.

---

5 The relative activity of a deformation mode is defined as its relative contribution to the volume averaged plastic strain tensor.

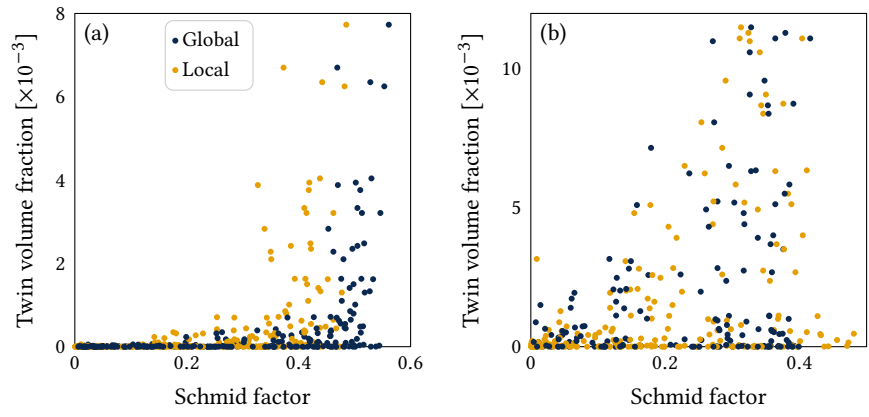

**Figure 5**  Twin volume fraction as a function of the local and global Schmid factors: (a) after yielding under uniaxial compression (−0.16 %); (b) at the end of a uniaxial compression test (−0.5 %).

**Table 2**  Average Schmid factor and corresponding standard deviation calculated at the end of the compression test for the most active twin variants (i.e. with a volume fraction higher than 1 %) with either the global or local definition of the Schmid factor.

|  | Local | Global |
| --- | --- | --- |
| Average | 0.36 | 0.38 |
| Standard deviation | 0.037 | 0.064 |

As illustrated by Figure 4, important intragranular internal stresses are produced as a result of twinning. Figure 6 shows the evolution of the equivalent von Mises stress within a twinned grain. According to the numerical results, important fluctuations of the stress field within a twinned grain are observed. Specifically, while local stress maxima are obtained near twin boundaries, local stress minima are located close to the center of twinned domains. The later result suggests that stress relaxation occurs within twins.

**Figure 6**  Distribution of the twin volume fraction and von Mises equivalent stress along a line path within a twinned grain. The line path is shown in Figure 4.

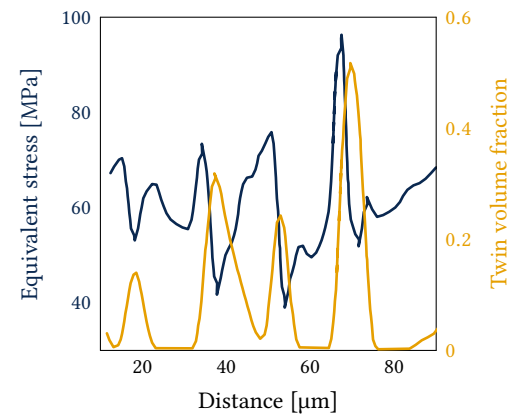

In the recent years, 3D XRD techniques have been used to investigate the evolution of internal stresses within twinned grains. For a given twin-parent pair, these techniques provide some estimates of the average stress tensor for each single grain. In the following, the average stress tensor of a parent grain (respectively child grain) is denoted by $\bar{\sigma}^m$ (respectively $\bar{\sigma}^c$). For the investigation of internal stresses, it is convenient to introduce the following stress quantities:

$$\bar{\sigma}^m = \boldsymbol{k}_c \cdot \bar{\sigma}^m \cdot \boldsymbol{k}_c \text{ and } \bar{\sigma}^c = \boldsymbol{k}_c \cdot \bar{\sigma}^c \cdot \boldsymbol{k}_c, \tag{31}$$

$$\bar{\tau}^m = \boldsymbol{t}_c \cdot \bar{\sigma}^m \cdot \boldsymbol{k}_c \text{ and } \bar{\tau}^c = \boldsymbol{t}_c \cdot \bar{\sigma}^c \cdot \boldsymbol{k}_c. \tag{32}$$

In the above equations, $\bar{\sigma}^m$ (respectively $\bar{\sigma}^c$) is the average normal stress exerted by the parent (respectively child) grain on the twin habit plane. In a similar fashion, $\bar{\tau}^m$ (respectively $\bar{\tau}^c$) is the average tangential stress exerted by the parent (respectively child) grain on the twin habit plane along the twinning direction.

For each twin-parent pair, the average normal stress $\bar{\sigma}^c$ in the twin is plotted as a function of that of the corresponding parent grain $\bar{\sigma}^m$ in Figure 7. The numerical results are in agreement with experimental results from literature (Abdolvand et al. 2015; Louca et al. 2021) showing that, for each pair, the average normal stresses within twins and parents are close to each other.

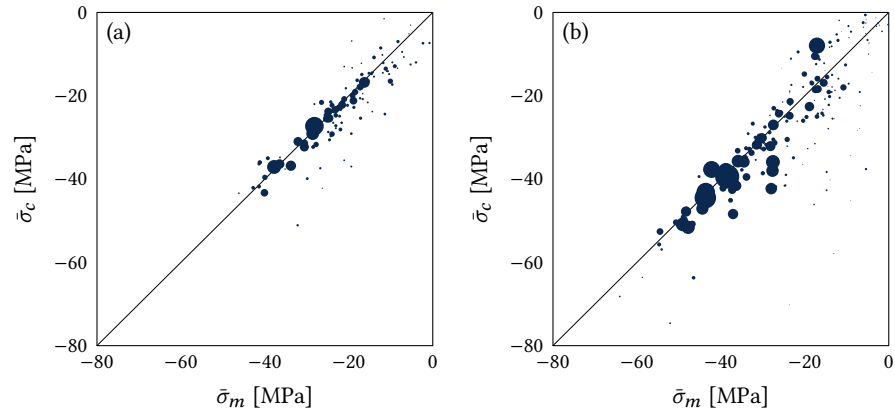

**Figure 7**  Average normal stress exerted by the child grain $\bar{\sigma}^c$ on the twin habit plane as a function of that of the corresponding parent grain $\bar{\sigma}^m$. Normal stresses evaluated after (a) yielding under uniaxial compression ($-0.16\%$) and (b) at the end of the uniaxial compression test ($-0.5\%$). Dot size proportional to the equivalent spherical radius of the child grain.

Louca et al. (2021) observed that, immediately after yielding, the tangential stress within a twin is in average higher than that of the corresponding parent grain. According to numerical results shown in Figure 8, this aspect is underestimated by the present model. The possible reason for such discrepancies might be the fact that the development of plastic strains is not affected by twin boundaries. Indeed, while twin boundaries may limit dislocation motion, their contribution to strain hardening is not considered here. Also, with further deformation under compression, Louca et al. (2021) found that large twins have either similar or lower tangential stresses than their parents. This effect, which is attributed to the opposition of surrounding grains to the accumulation of shear deformation within twinned domains, is correctly reproduced by the proposed model. Indeed, as shown in Figure 8, for large twins, the tangential stresses are significantly relaxed compared to those of the corresponding parent grains at the end of the compression test.

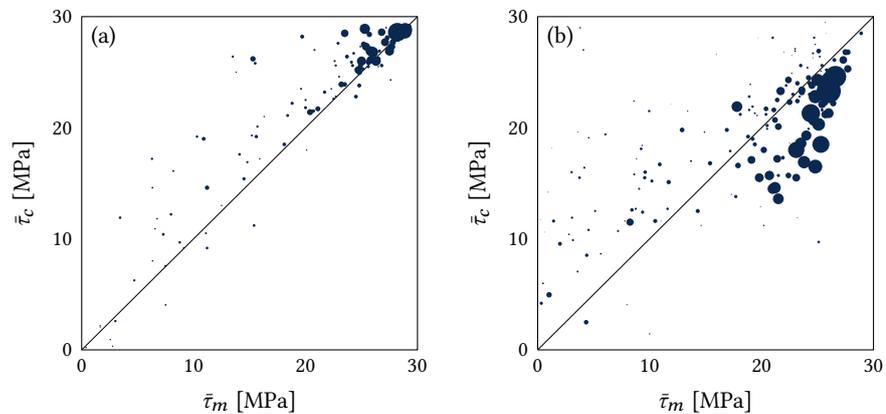

**Figure 8**  Average tangential stress exerted by the child grain $\bar{\tau}^c$ on the twin habit plane along the twinning direction as a function of that of the corresponding parent grain $\bar{\tau}^m$. Tangential stresses have been evaluated (a) after yielding under uniaxial compression ($-0.16\%$) and (b) at the end of the uniaxial compression test ($-0.5\%$). Dot size proportional to the equivalent spherical radius of the child grain.

### 3.4.2 Uniaxial compression and tension

The numerical and experimental stress-strain curves obtained for the compression-tension loading path along the rolling direction are shown in Figure 9. The deformation behavior under compression is correctly reproduced by the model. Specifically, experimental results indicate a strong Bauschinger effect since yielding occurs almost immediately after load reversal (around $-53$ MPa). According to the calculated deformation mode activities, see Figure 9, yielding after load reversal is the consequence of both the reactivation of basal slip systems and detwinning. The reactivation of basal systems is made possible thanks to the internal stresses generated during

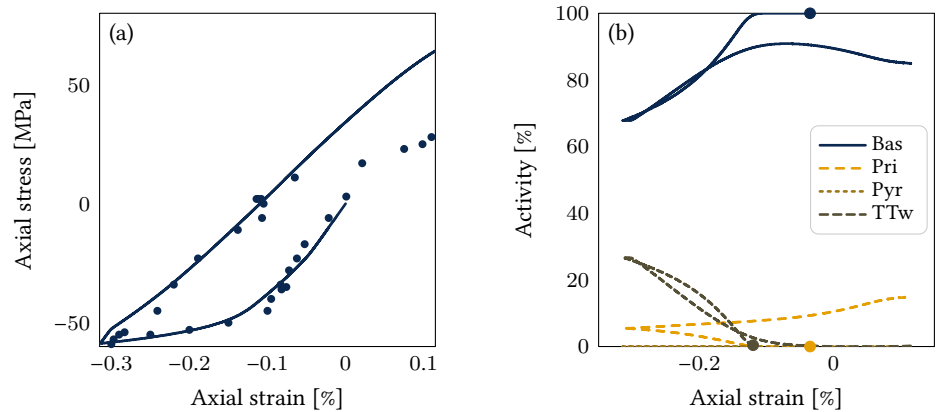

**Figure 9** (a) Experimental and numerical stress-strain curves obtained for a uniaxial compression and tension test along the rolling direction. Dots correspond to the experimental data of Louca et al. (2021) while solid lines correspond to numerical data. (b) Relative activities of the different inelastic deformation modes. The first activation of a given type of slip or twinning system is represented with a dot.

deformation under uniaxial compression. However, the internal stresses are not high enough to reactivate prismatic slip systems. Also, the activity of twinning systems tends toward zero when the axial load is reversed toward tension, which indicates that full detwinning has occurred.

Although basal and prismatic slip systems contribute to plastic deformation under uniaxial tension, the axial stress is overestimated by the proposed model. This poor description of the behavior under tension is likely due to the description of hardening, which relies on a purely isotropic formulation. While the internal stresses resulting from plastic strain incompatibilities contribute to kinematic hardening at the macroscopic scale, kinematic hardening is not considered at the microscale *via* constitutive relations. A more realistic description of the behavior under tension would therefore require considering this additional contribution. Specifically, in contrast with the adopted isotropic hardening rule, a kinematic hardening rule would facilitate the development of plasticity upon load reversal. As a result, when switching from uniaxial compression to uniaxial tension, the consideration of kinematic hardening would promote the development of plasticity, hence reducing the axial stress, which would be more consistent with the experimental results obtained after load reversal.

The basal pole figures obtained from numerical simulations with the above method are plotted in Figure 10. The alignment of basal poles from the normal direction toward the rolling direction at the end of the compression stage, which has been observed by Louca et al. (2021), is correctly depicted by the present model. Also, as a consequence of detwinning, basal poles are rotated back toward the normal direction of the specimen at the end of the tension stage.

## 4 Conclusions

A crystal plasticity-based constitutive model for simulating formation of twins in metallic polycrystals is developed. To include the contribution of twinning, each material point is treated as a composite material consisting of a parent constituent and multiple twin variants. The twin volume fractions and their spatial gradients are treated as external state variables to account for the contribution of twin boundaries to free energy. The resulting constitutive model is therefore non-local since the evolution of the twin volume fraction depends on the state of neighbouring material points.

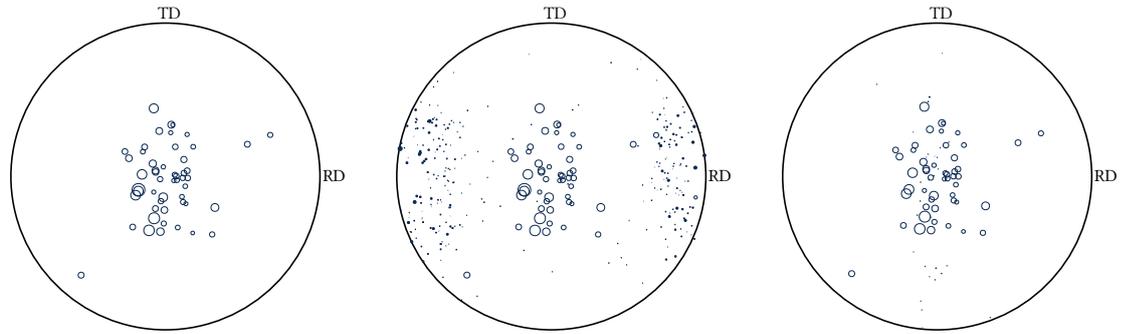

**Figure 10**    Basal pole figures obtained during a compression and tension test: (left) initial, (middle) intermediate and (right) final. Intermediate and final pole figures correspond to the end of the compression stage (−0.31 %) and the end of the tension stage (0.13 %), respectively. Dot size is proportional to the equivalent spherical radius of the grain. Empty circles correspond to parent grains while filled circles correspond to child grains.

The proposed set of constitutive relations has been implemented in a spectral solver to model the behavior of a AZ31 magnesium alloy. For the investigated loading conditions, both crystallographic slip and twinning contribute to macroscopic plastic deformation. Specifically, the initial yielding during uniaxial compression coincides with the activation of basal slip systems while the transition toward the plastic regime is attributed to twinning. When the loading direction is reversed toward tension, yielding occurs almost immediately after load reversal because of the joint contributions of detwinning and basal slip. Also, according to the numerical results, the development of texture and internal stresses is consistent with the experimental observations of the literature. Specifically, the proposed model naturally replicates the morphological texture evolution resulting from the concentration of plastic shear strains within twin domains.

Future work should focus on the role of twin boundaries on strain hardening. Indeed, while experimental studies indicate that twin boundaries contribute to strain hardening (Basinski et al. 1997), the impact of twin boundaries on the resistance to dislocation motion has been ignored in the present work. The coupling between slip and twinning deformation modes should be considered for future developments. For this purpose, the possibility of including the role of twin volume fraction gradients within the hardening rule will be explored. Indeed, twin boundaries correspond to the regions where such gradients are important. Twin volume fraction gradients should therefore affect the development of plasticity. Also, the interactions of twins with grain boundaries, which have been ignored here, should be included in constitutive relations to investigate the impact of grain size on the development of twins.

## A   Numerical implementation

The strategy used to compute the macroscopic behavior of a periodic volume element is presented in Algorithm A.1. It relies on an iterative procedure whose objective is twofold. First, when stress-controlled or mixed boundary conditions are prescribed to volume element, the iterative procedure searches for the macroscopic strain state that satisfies boundary conditions. For this purpose, it uses a convergence indicator $e_{bc}$ that is computed according to

$$e_{bc} = \frac{||\Sigma - \Sigma_{bc}||}{||\Sigma||} \tag{A.1}$$

where $\Sigma$ is the macroscopic stress tensor and $\Sigma_{bc}$ is the prescribed macroscopic stress tensor. Also, the iterative procedure aims at finding the strain tensor field that verifies compatibility, equilibrium and constitutive equations. The convergence indicator $e_{eq}$ that allows determining whether equilibrium is reached is computed according to

$$e_{eq} = \frac{\text{rms}(\boldsymbol{\sigma} \cdot \boldsymbol{\nabla})}{||\Sigma||} \tag{A.2}$$

```
 1  Set the state variables to their initial values
 2  Compute the modified Green tensor in frequency domain
 3  for each increment j of a deformation process do
 4      Compute time at the end of the increment
 5      j ← 1
 6      while e_bc > e_bc,c do
 7          Estimate the macroscopic strain and rotation tensors from boundary conditions
 8          while e_eq > e_eq,c do
 9              if i = 1 and j = 1 then
10                  Use the Voigt approximation to obtain initial estimates for the strain and rotation tensor fields
11              else
12                  Use the spectral method to obtain new estimates for the strain and rotation tensor fields
13              end if
14              for each voxel do
15                  for each child or parent constituent do
16                      Use the Sachs approximation to obtain the corresponding strain tensor
17                      Use the Taylor approximation to obtain the corresponding rotation tensor
18                      Integrate constitutive relations
19                  end for
20                  for each child constituent do
21                      Compute the divergence of the microstress vector
22                      Compute the twin volume fraction
23                  end for
24                  Compute the local stiffness tensor
25                  Compute the local plastic strain tensor
26              end for
27              Compute the convergence indicator e_eq
28              j ← j + 1
29          end while
30          Compute the macroscopic stress tensor
31          Compute the convergence indicator e_bc
32      end while
33      Save state variables
34      for Each voxel do
35          for Each child or parent constituent do
36              Update Euler angles
37          end for
38      end for
39  end for
```

**Algorithm A.1** Computation of the effective behavior of a volume element. The tolerances on the boundary conditions and static equilibrium conditions are denoted by $e_{bc,c}$ and $e_{eq,c}$, respectively.

where "rms" denotes the root mean square. In the present work, the fourth-order Runge-Kutta method is used for the integration of constitutive equations.

## B  Mesh dependence

To investigate the sensitivity to resolution, simulations of a uniaxial compression test with resolutions of $64^3$, $96^3$, $128^3$ and $192^3$ voxels were conducted.

The simulation conditions are the same as those given in Section 3. As shown in Figure B.1(a), the macroscopic behavior under uniaxial compression is correctly evaluated, even for a resolution of $64^3$.

A minimum resolution of $128^3$ is however needed to estimate the grain averaged stress/strain tensors. For such a resolution, the average axial stress or strain within a parent grain at the end of a uniaxial compression test is similar to that obtained for a resolution of $192^3$ voxels, see Figure B.1(b).

## References

Abdolvand, H., M. Daymond, and C. Mareau (2011). Incorporation of twinning into a crystal plas-

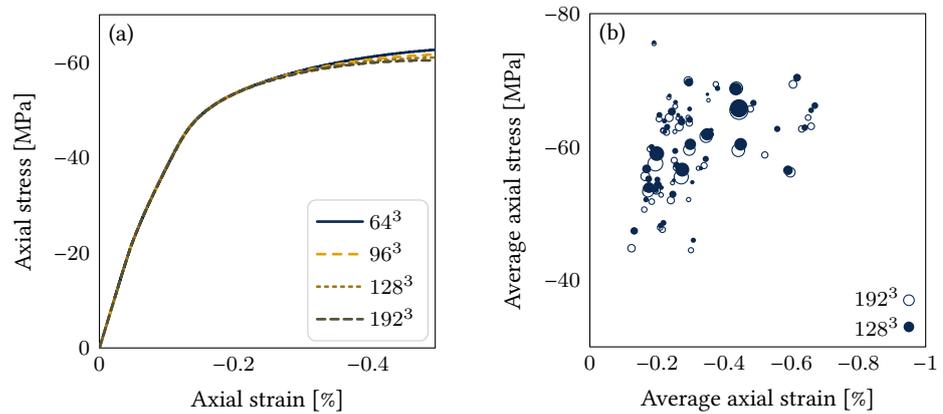

**Figure B.1** (a) Uniaxial compression macroscopic stress-strain curves. (b) Parent grain state at the end of a uniaxial compression test for two resolutions.

ticity finite element model: Evolution of lattice strains and texture in Zircaloy-2. *International Journal of Plasticity* 27(11):1721–1738. [DOI].

Abdolvand, H., M. Majkut, J. Oddershede, S. Schmidt, U. Lienert, B. Diak, P. Withers, and M. Daymond (2015). On the deformation twinning of Mg AZ31B: A three-dimensional synchrotron X-ray diffraction experiment and crystal plasticity finite element model. *International Journal of Plasticity* 70:77–97. [DOI].

Agnew, S., D. Brown, and C. Tomé (2006). Validating a polycrystal model for the elastoplastic response of magnesium alloy AZ31 using in situ neutron diffraction. *Acta Materialia* 54(18):4841–4852. [DOI].

Ardeljan, M., I. J. Beyerlein, B. A. McWilliams, and M. Knezevic (2016). Strain rate and temperature sensitive multi-level crystal plasticity model for large plastic deformation behavior: Application to AZ31 magnesium alloy. *International Journal of Plasticity* 83:90–109. [DOI].

Ardeljan, M., R. McCabe, I. Beyerlein, and M. Knezevic (2015). Explicit incorporation of deformation twins into crystal plasticity finite element models. *Computer Methods in Applied Mechanics and Engineering* 295(C):396–413. [DOI].

Arul Kumar, M., B. Clausen, L. Capolungo, R. McCabe, W. Liu, J. Tischler, and C. Tomé (2018). Deformation twinning and grain partitioning in a hexagonal close-packed magnesium alloy. *Nature Communications* 9:1–8. [DOI], [OA].

Aydıner, C., J. Bernier, B. Clausen, U. Lienert, C. Tomé, and D. Brown (2009). Evolution of stress in individual grains and twins in a magnesium alloy aggregate. *Physical Review B* 80(2):024113. [DOI], [HAL].

Basinski, Z., M. Szczerba, M. Niewczas, J. Embury, and S. Basinski (1997). The transformation of slip dislocations during twinning of copper-aluminum alloy crystals. *Revue de Métallurgie* 94(9):1037–1044. [DOI], [HAL].

Brown, D., S. Agnew, M. Bourke, T. Holden, S. Vogel, and C. Tomé (2005). Internal strain and texture evolution during deformation twinning in magnesium. *Materials Science and Engineering: A* 399(1):1–12. [DOI].

Bunge, H. (1968). The orientation distribution function of the crystallites in cold-rolled and annealed low-carbon steel sheets. *Physica Status Solidi (b)* 26(1):167–172. [DOI].

Cermelli, P. and M. Gurtin (2002). Geometrically necessary dislocations in viscoplastic single crystals and bicrystals undergoing small deformations. *International Journal of Solids and Structures* 39:6281–6309. [DOI].

Chang, Y. and D. Kochmann (2015). A variational constitutive model for slip-twinning interactions in hcp metals: Application to single- and polycrystalline magnesium. *International Journal of Plasticity* 73:39–61. [DOI], [OA].

Cheng, J. and S. Ghosh (2015). A crystal plasticity FE model for deformation with twin nucleation in magnesium alloys. *International Journal of Plasticity* 67:148–170. [DOI].

Cheng, J. and S. Ghosh (2017). Crystal plasticity finite element modeling of discrete twin evolution in polycrystalline magnesium. *Journal of the Mechanics and Physics of Solids* 99(C):512–538.

[DOI], [HAL].

Cherkaoui, M. (2003). Constitutive equations for twinning and slip in low-stacking-fault-energy metals: a crystal plasticity-type model for moderate strains. *Philosophical Magazine* 83(31-34):3945–3958. [DOI].

Clausen, B., C. Tomé, D. Brown, and S. Agnew (2008). Reorientation and stress relaxation due to twinning: Modeling and experimental characterization for Mg. *Acta Materialia* 56(11):2456–2468. [DOI].

Clayton, J. and J. Knap (2011). A phase field model of deformation twinning: Nonlinear theory and numerical simulations. *Physica D: Nonlinear Phenomena* 240(9):841–858. [DOI], [OA].

Eisenlohr, P., M. Diehl, R. Lebensohn, and F. Roters (2013). A spectral method solution to crystal elasto-viscoplasticity at finite strains. *International Journal of Plasticity* 46:37–53. [DOI].

Frémond, M. and B. Nedjar (1996). Damage, gradient of damage and principle of virtual power. *International Journal of Solids and Structures* 33(8):1083–1103. [DOI].

Grilli, N., A. Cocks, and E. Tarleton (2020a). A phase field model for the growth and characteristic thickness of deformation-induced twins. *Journal of the Mechanics and Physics of Solids* 143:104061. [DOI], [OA].

Grilli, N., E. Tarleton, P. Edmondson, M. Gussev, and A. Cocks (2020b). In situ measurement and modelling of the growth and length scale of twins in $\alpha$-uranium. *Physical Review Materials* 4:043605. [DOI], [OA].

Guo, Y., H. Abdolvand, T. Britton, and A. Wilkinson (2017). Growth of $\{11\bar{2}2\}$ twins in titanium: A combined experimental and modelling investigation of the local state of deformation. *Acta Materialia* 126:221–235. [DOI], [OA].

Gurtin, M., E. Fried, and L. Anand (2010). *The Mechanics and Thermodynamics of Continua*. Cambridge University Press. [DOI].

Gussev, M., P. Edmondson, and K. Leonard (2018). Beam current effect as a potential challenge in SEM-EBSD in situ tensile testing. *Materials Characterization* 146:25–34. [DOI].

Indurkar, P., S. Baweja, R. Perez, and S. Joshi (2020). Predicting textural variability effects in the anisotropic plasticity and stability of hexagonal metals: Application to magnesium and its alloys. *International Journal of Plasticity* 132:102762. [DOI].

Jiang, L., J. Jonas, R. Mishra, A. Luo, A. Sachdev, and S. Godet (2007). Twinning and texture development in two Mg alloys subjected to loading along three different strain paths. *Acta Materialia* 55(11):3899–3910. [DOI].

Juan, P.-A., S. Berbenni, M. Barnett, C. Tomé, and L. Capolungo (2014). A double inclusion homogenization scheme for polycrystals with hierarchal topologies: application to twinning in Mg alloys. *International Journal of Plasticity* 60:182–196. [DOI].

Kalidindi, S. R. (2001). Modeling anisotropic strain hardening and deformation textures in low stacking fault energy fcc metals. *International Journal of Plasticity* 17(6):837–860. [DOI].

Kondo, R., Y. Tadano, and K. Shizawa (2014). A phase-field model of twinning and detwinning coupled with dislocation-based crystal plasticity for HCP metals. *Computational Materials Science* 95:672–683. [DOI].

Liu, C., P. Shanthraj, M. Diehl, F. Roters, S. Dong, J. Dong, W. Ding, and D. Raabe (2018). An integrated crystal plasticity–phase field model for spatially resolved twin nucleation, propagation, and growth in hexagonal materials. *International Journal of Plasticity* 106:203–227. [DOI].

Louca, K., H. Abdolvand, C. Mareau, M. Majkut, and J. Wright (2021). Formation and annihilation of stressed deformation twins in magnesium. *Communications Materials* 2(9):1–11. [DOI], [OA].

Mareau, C. and M. Daymond (2011). Comparison of experimentally determined texture development in Zircaloy-2 with predictions from a rate-dependent polycrystalline model. *Materials Science and Engineering: A* 528(29):8676–8686. [DOI], [HAL].

Mareau, C. and M. Daymond (2016). Micromechanical modelling of twinning in polycrystalline materials: Application to magnesium. *International Journal of Plasticity* 85:156–171. [DOI], [HAL].

Moulinec, H. and P. Suquet (1998). A numerical method for computing the overall response of nonlinear composites with complex microstructure. *Computer Methods in Applied Mechanics* and Engineering 157(1):69–94. [DOI], [ARXIV].

Nervo, L., A. King, A. Fitzner, W. Ludwig, and M. Preuss (2016). A study of deformation twinning in a titanium alloy by X-ray diffraction contrast tomography. *Acta Materialia* 105:417–428. [DOI], [OA].

Nguyen, Q.-S. (2000). *Stability and Nonlinear Solid Mechanics*. Wiley. ISBN: 9780471492887.

Nguyen, Q.-S. (2015). Some remarks on standard gradient models and gradient plasticity. *Mathematics and Mechanics of Solids* 20(6):760–769. [DOI], [HAL].

Paramatmuni, C. and F. Dunne (2020). Effect of twin crystallographic orientation on deformation and growth in Mg alloy AZ31. *International Journal of Plasticity* 135:102775. [DOI].

Paramatmuni, C., Y. Guo, P. Withers, and F. Dunne (2021). A three-dimensional mechanistic study of the drivers of classical twin nucleation and variant selection in Mg alloys: a mesoscale modelling and experimental study. *International Journal of Plasticity* 143:103027. [DOI].

Paramatmuni, C. and A. Kanjarla (2019). A crystal plasticity FFT based study of deformation twinning, anisotropy and micromechanics in HCP materials: Application to AZ31 alloy. *International Journal of Plasticity* 113:269–290. [DOI].

Pei, Z., X. Zhang, T. Hickel, M. Friák, S. Sandlöbes, B. Dutta, and J. Neugebauer (2017). Atomic structures of twin boundaries in hexagonal close-packed metallic crystals with particular focus on Mg. *npj Computational Materials* 3(6). [DOI], [OA].

Poulsen, H., S. Nielsen, E. Lauridsen, S. Schmidt, R. Suter, U. Lienert, L. Margulies, T. Lorentzen, and D. Juul Jensen (2001). Three-dimensional maps of grain boundaries and the stress state of individual grains in polycrystals and powders. *Journal of Applied Crystallography* 34(6):751–756. [DOI], [HAL].

Proust, G., C. Tomé, and G. Kaschner (2007). Modeling texture, twinning and hardening evolution during deformation of hexagonal materials. *Acta Materialia* 55(6):2137–2148. [DOI].

Ravaji, B. and S. Joshi (2021). A crystal plasticity investigation of grain size-texture interaction in magnesium alloys. *Acta Materialia* 208:116743. [DOI], [HAL].

Simmons, G. and H. Wang (1971). *Single Crystal Elastic Constants and Calculated Aggregate Properties: A Handbook*. The MIT Press. ISBN: 9780262190923.

Staroselsky, A. and L. Anand (2003). A constitutive model for hcp materials deforming by slip and twinning: application to magnesium alloy AZ31B. *International Journal of Plasticity* 19(10):1843–1864. [DOI].

Wang, H., P. Wu, S. Kurukuri, M. Worswick, Y. Peng, D. Tang, and D. Li (2018). Strain rate sensitivities of deformation mechanisms in magnesium alloys. *International Journal of Plasticity* 107:207–222. [DOI], [OA].

Wu, L., S. Agnew, D. Brown, G. Stoica, B. Clausen, A. Jain, D. Fielden, and P. Liaw (2008). Internal stress relaxation and load redistribution during the twinning–detwinning-dominated cyclic deformation of a wrought magnesium alloy, ZK60A. *Acta Materialia* 56(14):3699–3707. [DOI].

Xu, F., R. Holt, and M. Daymond (2008a). Modeling lattice strain evolution during uniaxial deformation of textured Zircaloy-2. *Acta Materialia* 56(14):3672–3687. [DOI].

Xu, F., R. Holt, and M. Daymond (2009). Modeling texture evolution during uni-axial deformation of Zircaloy-2. *Journal of Nuclear Materials* 394(1):9–19. [DOI].

Xu, F., R. Holt, M. Daymond, R. Rogge, and E. Oliver (2008b). Development of internal strains in textured Zircaloy-2 during uni-axial deformation. *Materials Science and Engineering: A* 488(1):172–185. [DOI].

Zhang, J. and S. Joshi (2012). Phenomenological crystal plasticity modeling and detailed micromechanical investigations of pure magnesium. *Journal of the Mechanics and Physics of Solids* 60:945–972. [DOI].

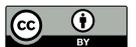

**Authors' contributions**     C.M. developed constitutive equations, codes and conducted numerical simulations. C.M. and H.A. interpreted numerical data. C.M. wrote the first draft. C.M. and H.A. contributed to the final version of the manuscript.

**Supplementary Material**     Figure data in text format. See permalink 10.5281/zenodo.6342152.

**Acknowledgements**     None.

**Ethics approval and consent to participate**     Not applicable.

**Consent for publication**     Not applicable.

**Competing interests**     The authors declare that they have no competing interests.

**Journal's Note**     JTCAM remains neutral with regard to the content of the publication and institutional affiliations.